\documentclass[useAMS,onecolumn]{mn2e}

\def\be{\begin{equation}}
\def\ee{\end{equation}}
\def\bea{\begin{eqnarray}}
\def\eea{\end{eqnarray}}
\def\etal{et al.~}
\def\ge{\gamma_e}
\def\gemin{\gamma_{e,{\rm min}}}
\def\gm{\gamma_m}

\title[]{High Energy Afterglow Emission from Gamma-Ray Bursts}
\author[]{Yi-Zhong Fan$^{1,2,4}$\thanks{Golda Meir Fellow, E-mail: yzfan@pmo.ac.cn}, Tsvi Piran$^1$
\thanks{tsvi@phys.huji.ac.il}, Ramesh Narayan$^3$\thanks{narayan@cfa.harvard.edu}
and Da-Ming Wei$^{2}$\thanks{dmwei@pmo.ac.cn}\\
$^1${\sl The Racah Inst. of Physics, Hebrew University, Jerusalem 91904, Israel}\\
$^2$ {\sl Purple Mountain Observatory, Chinese Academy of Science,
Nanjing 210008, China}\\
$^3$ {\sl Institute for Theory and Computation, Center for
  Astrophysics, Harvard University, 60 Garden St., Cambridge, MA,
  02138}\\
 $^4${\sl Neils Bohr International Academy, Niels Bohr Institute,
 University of Copenhagen, Blegdamsvej 17, 2100 Copenhagen, Denmark} }
\date{Accepted ......
Received ......; in original form ......}

\begin{document}

\maketitle
\begin{abstract}
We calculate the very high energy (sub-GeV to TeV) inverse Compton
emission of GRB afterglows. We argue that this emission provides a
powerful test of the currently accepted afterglow model. We focus on
two processes: synchrotron self-Compton (SSC) emission within the
afterglow blast wave, and external inverse Compton (EIC) emission
which occurs when flare photons (produced by an internal process)
pass through the blast wave. We show that if our current
interpretations of the Swift XRT data are correct, there should be a
canonical high energy afterglow emission light curve. Our
predictions can be tested with high energy observatories such as
GLAST, Whipple, H.E.S.S. and MAGIC. Under favorable conditions we
expect afterglow detections in all these detectors.

\end{abstract}



\section{Introduction}
\label{sec:SSC1}

EGRET detected more than 30 Gamma-Ray Bursts (GRBs) with GeV
emission (Schneid et al. 1992; Sommer et al. 1994; Hurley et al.
1994; Schaefer et al. 1998; Gonz\'alez et al. 2003).  The highest
energy photon detected was the 18 GeV photon which arrived 4500
seconds after the trigger of GRB 940217 (Hurley et al. 1994). These
observations motivated many interesting ideas. Some focused on
prompt high energy photon emission, e.g., synchrotron-self-Compton
(SSC) emission or inverse Compton scattering of photons emitted by
one shell by electrons in another shell \cite{TK05}. Others focused
on high energy afterglow processes: the interaction of
ultra-relativistic protons with a dense cloud (Katz 1994), SSC in
early forward and reverse shocks (M\'esz\'aros \& Rees 1994),
electromagnetic cascade of TeV $\gamma-$rays in the
infrared/microwave background (Plaga 1995), synchrotron radiation of
ultra-high energy forward shock protons \cite{TT98}, and inverse
Compton scattering of prompt $\gamma-$rays by reverse shock
electrons (Beloborodov 2005).

Two kinds of high energy afterglow emission models have been discussed
extensively. The first is SSC emission. Motivated by the successful
detection of an optical flash in GRB 990123 (Akerlof et al.  1999;
Sari \& Piran 1999; M\'esz\'aros \& Rees 1999), Wang, Dai \& Lu
(2001a, b), Pe'er \& Waxman (2005) and Kobayashi \etal (2007)
calculated SSC emission from the reverse shock.  Granot \& Guetta
(2003) and Pe'er \& Waxman (2004) applied these ideas to GRB
941017. The high energy SSC component of the forward component was
calculated by Dermer, Chiang \& Mitman (2000), Sari \& Esin (2001),
and Zhang \& M\'esz\'aros (2001b). The second family of models
involves the external inverse Compton (EIC) process.  These include
Comptonization of the prompt photons by the forward shock electrons
(Fan, Zhang \& Wei 2005b), and upscattering of far-UV/X-ray flare
photons (assuming that they originate in internal shocks) by the
forward shock (Wang, Li \& M\'esz\'aros 2006; Fan \& Piran 2006b).

Most of the above calculations were based on the standard afterglow
model.  However, recently, {\it Swift} has detected numerous GRBs
whose early (first $10^4$ s) afterglow emission cannot be reproduced
within the standard model \cite{M06,PF06,Z07}. Various modifications
of the standard model have been put forward to explain the
observations. However, none are compelling and the validity of the
whole model is now in question.

High energy emission provides a new window into afterglow physics and
can provide an independent test of models.  Motivated by this, we
calculate the predicted high energy afterglow emission in different
scenarios. We show that there is a canonical high energy GRB afterglow
light curve which ought to be observed (see Fig.
\ref{fig:summary}). The detection of the predicted high energy
emission features by observations with GLAST or ground-based gamma-ray
detectors would enable us to test the validity of the overall model as
well as the specific modifications that have been put forward to
explain {\it Swift} observations.

The paper is structured as follows. In section 2 we review {\it
Swift} GRB afterglow observations and their interpretation. In
section \ref{sec:code} we describe the methods we employ for
careful calculations of the inverse Compton effect; this section
may be skipped if one is interested only in the results. In
section 4 we calculate the SSC emission of the forward shock, and
in section 5 we calculate the possible high energy emission
associated with X-ray flares, including both SSC emission from
within the flare and EIC emission from the forward shock. In
section \ref{sec:det} we discuss the prospects for detecting high
energy afterglows by GLAST and ground-based telescopes.  We
conclude in section 7 with a summary.

\section{{\it Swift} GRB afterglow observations}\label{sec:Review}

\begin{figure}
\begin{picture}(0,200)
\put(0,0){\includegraphics{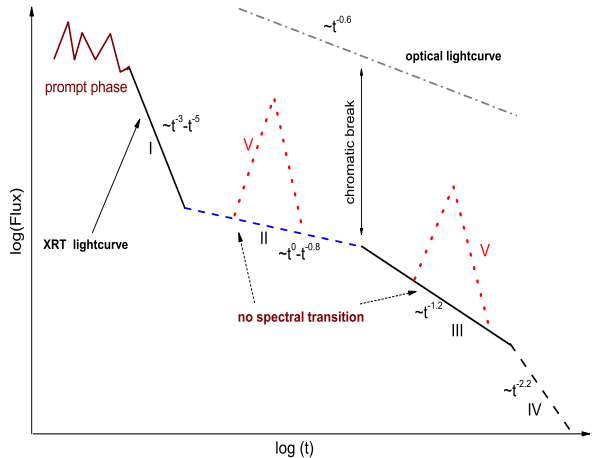}}
\end{picture}
\caption{Schematic cartoon of the X-ray light curve of a GRB and its
afterglow, based on {\it Swift} XRT data (see Zhang et al. 2006 and
Nousek et al. 2006 for similar plots).  Also shown is a schematic
optical light curve, which often does not show the same breaks as the
X-ray light curve (Fan \& Piran 2006a; Panaitescu \etal 2006; Huang et
al. 2007).} \label{fig:Cartoon}
\end{figure}

In the pre-{\it Swift} era, most of the afterglow data was collected
hours after the GRB. These data were found to be consistent with the
external forward shock model, though sometimes energy injection, a
wind profile, or structured or patchy jets had to be invoked to
account for the observations (Piran 2004).  The {\it Swift} satellite
has changed the situation.  The X-ray Telescope (XRT) and the
UV/optical Telescope (UVOT) onboard this satellite can slew to the
direction of a GRB in real time and record the early broad band
afterglow light curves. A schematic X-ray afterglow light curve based
on the XRT data has been summarized by Zhang et al. (2006) and Nousek
et al. (2006) (see Fig. \ref{fig:Cartoon}) and consists of the
following features: A very early sharp decline (phase-I); A shallow
decline of the X-ray afterglow (phase-II); A ``normal" decay phase
(phase-III), possibly followed by a jet break (phase-IV); Energetic
X-ray flares (phase-V), which may show up during any phase. Note that
not all of these features have been detected in every burst. We focus
here on the most remarkable of the new features: the slow decline
(phase-II) and the flares (phase-V). Both are expected to have
associated signatures in the high energy emission.

In about half of the \emph{Swift} GRBs, the X-ray lightcurves show an
extended flattening (phase-II). In most cases, but not all, there is
no change in the spectral slope when the light curve makes a
transition from the shallow phase-II segment to the ``normal"
phase-III segment.  The usual interpretation of the shallow phase is
that it involves energy injection into the blast wave
\cite{Zhang06,Nousek06,GK06}. An alternative possibility is that the
parameter $\epsilon_e$, which measures the fraction of shock energy
transferred to the downstream electrons, varies with
time\footnote{Note, however, that for some GRBs the break in the X-ray
light curve is not accompanied by a break in the optical light curve
(see Figure \ref{fig:Cartoon}).  The interpretation of this chromatic
behavior is less clear. In the present work, we focus on the cases in
which the X-ray and optical light curves break achromatically.}, as
would be the case if this parameter is shock-strength dependent. In
either case the corresponding SSC emission of the forward shock would
be different from the one anticipated in the standard afterglow model,
as well as from each other.

Energetic X-ray flares (phase-V) have been detected in several
pre-{\it Swift} GRBs and in about half the {\it Swift} GRBs
\cite{Piro05,Burrows05,Galli06a,CG07}. The rapid decline of the
flares suggests that they arise due to ``late internal shocks"
resulting from reactivation of the central engine
\cite{FW05,Zhang06,Fan05a,King05,Nousek06,Falcone06,Perna06,PZ06,Zou06,Wu06,Dai06,GF06,LP07,Krimm07,CG07}.
An alternative interpretation is that the X-ray flares arise due
to refreshed shocks \cite{Piro05,Galli06a,Wu06,Guetta07}. Once
again, the GeV emission can serve to distinguish between the
models.

If the flares are produced by internal shocks, most of the
up-scattered photons would arrive after the far-UV/X-ray flare.
The high energy photons in this scenario will be produced by
scattering of the flare photons in the external shock. In the EIC
process, the duration of the high energy emission is stretched by
the spherical curvature of the blast wave (Beloborodov 2005; Wang
et al. 2006; Fan \& Piran 2006b) and is further extended by the
highly anisotropic distribution of the up-scattered photons (Fan
\& Piran 2006b; see our Fig.\ref{fig:WW} for a comparison). For
the latter effect,  see Aharonian \& Atoyan (1981), Ghisellini et
al. (1991) and Brunetti (2001) for details.

\section{Self Consistent Computation of Inverse Compton scattering
with Klein-Nishina suppression}\label{sec:code}

The relativistic electrons that are present in any synchrotron
source will also produce very high energy photons via inverse
Compton scattering (either SSC or EIC). We turn now to a
calculation of this emission. When the energy of the electrons and
the seed photons is sufficiently large it is necessary to take
into account the Klein-Nishina correction to the scattering cross
section.  We also need to include the inverse Compton cooling in
calculating the energy distribution of the relativistic electrons.

The essential problem is to calculate carefully the Compton
parameter, $Y$, the ratio between the power loss through inverse
Compton scattering and synchroton radiation ($P'_{\rm
ic}(\gamma_e)$ and $P'_{\rm s}(\gamma_e)$ respectively):
\begin{equation}
Y(\gamma_e)\equiv P'_{\rm ic}(\gamma_e)/P'_{\rm s}(\gamma_e).
\label{eq:Y_gamma}
\end{equation}
Throughout this work, the superscript $'$ indicates that the quantity
is measured in the rest frame of the emitting region.
In the regime of Thomson scattering, $Y$ is a constant, independent of
the electron Lorentz factor $\gamma_e$, and one obtains a constant
reduction in the amplitude of the synchrotron emission compared to the
case with no inverse Compton scattering.  This makes computations
relatively easy.  However, in the general case, since the
Klein-Nishina correction to $Y$ depends on $\gamma_e$, the effect of
inverse Compton scattering on the spectrum and on the electron energy
distribution is non-trivial.

The power emitted in synchrotron radiation by an electron with
Lorentz factor $\gamma_e$ is:
\begin{equation}
P'_{\rm s}(\gamma_e)=(\gamma_e^2-1)\sigma_T B'^2c/(6\pi),
\end{equation}
where $B'$ is the strength of the magnetic field. The corresponding
spectral energy distribution of the radiation is
\begin{equation}
P'_{\rm s}(\nu',\ge) d\nu' = P'_{\rm s}(\ge)
\,F\left[{\nu'\over\nu'_{\rm s}(\ge)}\right] {d\nu'\over \nu'_{\rm
s}(\ge)},
\end{equation}
where $\nu'_{\rm s}(\ge)=3\ge^2 e B'/(4\pi m_e c)$,
\begin{equation} F(x)=x\int^\infty_x K_{5/3}(\zeta)d\zeta,
 \end{equation}
and $K_{5/3}(\zeta)$ is the modified Bessel function.

The power emitted via inverse Compton scattering is given by:
\begin{equation}
P'_{\rm ic}(\gamma_e) =\int^\infty_0 h \nu'_{_{\rm ic}}{dN'_\gamma
\over dt d\nu'_{_{\rm ic}}} d\nu'_{_{\rm ic}} , \label{eq:P_compt}
\end{equation}
where $\nu'_{_{\rm ic}}$ is the frequency of the photon after
scattering.  The quantity $dN'_\gamma /dt d\nu'_{_{\rm ic}}$ is the
scattered photon spectrum per electron (Blumenthal \& Gould 1970).  It
is related to the spectral energy distribution of the inverse Compton
radiation emitted by an electron:
\begin{equation}
P'_{\rm ic}(\nu'_{\rm ic},\ge) d\nu'_{\rm ic} =h \nu'_{_{\rm
ic}}{dN'_\gamma \over dt d\nu'_{_{\rm ic}}} d\nu'_{\rm ic}.
\label{eq:Ramesh1}
\end{equation}

We define the auxiliary quantities $g \equiv \gamma_e h\nu'/(m_e
c^2)$, $f\equiv h\nu'_{_{\rm ic}}/(\gamma_e m_e c^2)$ and $q\equiv
f/[4g(1-f)]$, where $ h\nu$ is the photon energy before scattering.
The factor $g$ determines the regime of scattering, with the Thomson
limit corresponding to $g \ll 1$. The factor $f$ satisfies
$h\nu'/(\gamma_e m_e c^2) \leq f \leq 4g/(1+ 4g)$ (Jones 1968;
Blumenthal \& Gould 1970). We can express $dN'_\gamma /dt
d\nu'_{_{\rm ic}}$ in terms of these quantities and in terms of the
frequency distribution of the seed photons $n_{\nu'}$:
\begin{eqnarray}
{dN'_\gamma \over  dt d \nu'_{_{\rm ic}}} = {{3\sigma_T c \over
4\gamma_e^2}{n_{\nu'} d\nu' \over \nu'} [2q\ln q +(1+2q)(1-q)}
+{1\over 2}{(4g q)^2 \over 1+4g q}(1-q)]. \label{eq:Jones1}
\end{eqnarray}

\begin{figure}
\begin{picture}(0,200)
\put(0,0){\includegraphics{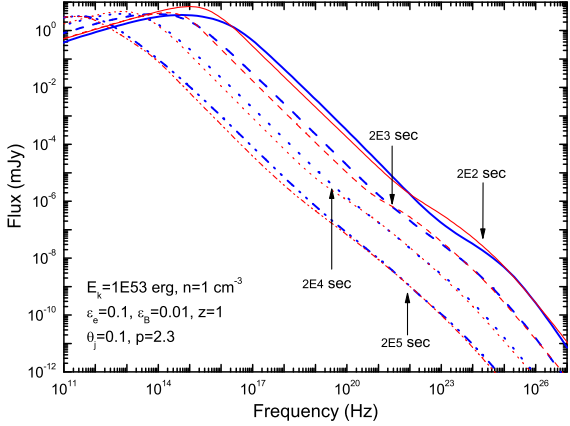}}
\end{picture}
\caption{Comparison of afterglow spectra calculated with the
instantaneous approximation (sec. 3.1, thin red lines) and the
dynamical approach (sec. 3.2, thick blue lines). Model parameters are
listed on the plot, and the timescales corresponding to each set of
spectra are identified next to the curves.  Note that all the results
shown in later plots were obtained with the full dynamical approach.
}
\label{fig:Ramesh_Yizhong}
\end{figure}

To complete the calculation we need to know the frequency
distribution of seed photons $n_{\nu'}$. For EIC this is simple
since the photons originate from an external source. For SSC,
however, the situation is more complicated.  This is because the
photons are produced via synchrotron emission by the same
electrons that are participating in inverse Compton scattering.
The additional cooling of these electrons by IC influences their
energy distribution and thus their synchrotron emission. We have
solved this problem by two different approaches. First, we have
used a simple ``instantaneous'' approach which involves a single
integral equation.  This method, which we describe in section 3.1,
is conceptually simple and computationally fast.  It is, however,
approximate.  We then describe in section 3.2 a more detailed and
general dynamical approach.  This more accurate method is the one
we have used for all the calculations presented later in this
paper. However, the two methods give very similar results in a
very wide energy range, as seen in Fig.\ref{fig:Ramesh_Yizhong}.

\subsection{Instantaneous  approximation}

In this approach we assume a functional form for the electron energy
distribution $n(\ge)$ produced through acceleration in the shock
front, and consider its instantaneous modification due to cooling.  An
electron of Lorentz factor $\ge$ has a cooling time given by
\begin{equation}
t'_c(\ge) = {\ge m_e c^2 \over P'_s(\ge)+P'_{\rm ic}(\ge)}.
\end{equation}
If $t'_c(\ge)$ is longer than the dynamical time $t'_d \sim R/{\Gamma
c}$, where $R$ is the radius of the shock front relative to the
central engine and $\Gamma$ is the bulk Lorentz factor of the outflow,
then the electron produces synchrotron and IC emission for the entire
time $t'_d$.  However, when $t'_c(\ge)$ is shorter than $t'_d$, the
electron radiates only for a time $t'_c(\ge)$. Thus, the total
spectral radiation density produced by all the electrons in the fluid
is given by
\begin{eqnarray}
\nonumber U_{\nu'}  \equiv n_{\nu'} h \nu' = \int_{\gemin}^\infty
\left[P'_{\rm s}(\nu',\ge) + P'_{\rm ic}(\nu',\ge)\right] \,
 \times  {\rm Min}[t'_d,t'_c(\ge)]\,
n(\ge)d\ge , \label{inteq}
\end{eqnarray}
The spectral power distributions $P'_{\rm s}(\nu',\ge)$ and $P'_{\rm
s}(\nu',\ge)$ are calculated as described earlier. For the inverse
Compton power, we write eq. (\ref{eq:Ramesh1}) as $P'_{\rm
ic}(\nu'_{\rm ic},\gamma_e) d\nu'_{\rm ic}\approx (1+g) c U_{\nu'}
\sigma(\nu',\gamma_e)d\nu'_{\rm ic}$, where $\sigma(\nu', \gamma_e)$
is the Klein-Nishina cross-section, which is equal to
\begin{eqnarray}
\sigma(\nu', \gamma_e) = {3\over4} \sigma_T \{{(1+g)\over g^3}
\left[ {2g(1+g)\over (1+2g)} -\ln(1+2g)\right] +
{1\over2g}\ln(1+2g)-{(1+3g)\over (1+2g)^2} \}.
\end{eqnarray}

Equation (\ref{inteq}) is an integral equation, since the function
$P'_{\rm ic}(\nu',\ge)$ inside the integral itself depends on
$U_{\nu'}$.  The quantity $\gemin$ is the smallest $\ge$ down to
which electrons are present. In dealing with equation (\ref{inteq})
we need to consider two cases (see Sari, Piran \& Narayan 1998 for
details and for the definitions of quantities):

\medskip\noindent{\bf Slow Cooling}: In this case, electrons with
$\ge=\gm$ have a cooling time $t'_c(\gm) > t'_d$. Then, $\gemin=\gm$,
and we may use equation (\ref{inteq}) directly with $\gemin=\gamma_m$
and $n(\gamma_e)$ given by the original energy distribution produced
in the shock.

\medskip\noindent{\bf Fast Cooling}: Here, all electrons with $\ge \geq
\gm$ have $t'_c(\ge) < t'_d$.  Therefore, electrons will continue
to cool below $\gm$ to a minimum $\gemin$ such that
\begin{equation}
t'_c(\gemin) = t'_d.
\end{equation}
Now, for the range $\gamma_{e,\rm min} \le \gamma_e < \gamma_m$, all
the electrons are available for radiating. Initially, most of the
electrons are at $\gamma_m$, and as these electrons cool
each electron will pass every $\gamma_e$ between
$\gamma_m$ and $\gamma_{e,\rm min}$ (where all these electrons accumulate).
Hence we have
\begin{equation}
n(\ge) \sim n(\gm), \qquad \gemin \leq \ge < \gm.
\end{equation}

As usual, we assume a power-law distribution for the electron
Lorentz factor:
\begin{equation}
n(\ge)d\ge \propto \ge^{-p} d\ge, \qquad \ge \geq \gm,
\end{equation}
for which $\gm$ is given by (Sari et al. 1998)
\begin{equation}
\gm = \epsilon_e \left({p-2\over p-1}\right) {m_p\over m_e}\,
(\Gamma-1)+1.
\end{equation}

Equation (\ref{inteq}) may be solved numerically via an iterative
method.  The algorithm proceeds as follows.  We begin with some
reasonable initial approximation for $U_{\nu'}$. Using this, we
compute $P'_{\rm ic}(\ge)$, $t'_c(\ge)$ and $\gemin$. Then, we compute
the spectral distributions $P'_s(\nu',\ge)$ and $P'_{\rm
ic}(\nu',\ge)$ for all $\ge \geq \gemin$ and obtain via equation
(\ref{inteq}) a new approximation for $U_{\nu'}$. We take this
$U_{\nu'}$, or (for smoother convergence) a suitable linear
combination of the new and old $U_{\nu'}$, as the current
approximation for $U_{\nu'}$ and repeat the steps. The iteration
usually converges fairly quickly.

This approach can be combined with any desired model for the GRB
fireball and afterglow dynamics.  We have used the dynamics described
in Sari \etal (1998), except that we multiplied the calculated fluxes
by a factor of 1/4 (cf., Yost et al. 2003).

\subsection{Dynamical approach}

In this approach we follow dynamically the electron distribution as
a function of time \cite{Moderski00}. The main uncertainty is from
the approximation for the initial distribution of the newly shocked
electrons as a function of time. Lacking a better model, we assume
that the electrons are accelerated at the shock wave initially to a
single power law distribution:
\begin{equation}
Q=K \gamma_e^{-p}$ for $\gamma_m \leq \gamma_e \leq \gamma_M,
\end{equation}
where the maximal Lorentz factor is given by $\gamma_M\approx
4\times 10^7 {B'}^{-1/2}$ (Wei \& Cheng 1997). The normalization
factor satisfies: $K \approx 4\pi (p-1) R^2 n_{\rm m}
\gamma_m^{p-1}$, and $n_{\rm m}$ is the number density of the
medium.
 We now follow the
evolution of the electron distribution using:
\begin{equation}
{\partial N_{\gamma_e} \over \partial R}+{\partial \over
\partial \gamma_e} (N_{\gamma_e} {d\gamma_e \over dR})=Q,
\label{eq:Fan0}
\end{equation}
where
\begin{equation}
{d \gamma_e \over dR}=-{\sigma_{\rm T} \over 6 \pi m_{\rm e}
c^2}{B'^2 \over \beta_\Gamma
\Gamma}[1+Y(\gamma_e)]\gamma_e^2-{\gamma_e \over R}, \label{eq:Fan1}
\end{equation}
$\Gamma$ is the bulk Lorentz factor of the shocked medium, and
$\beta_\Gamma=\sqrt{1-1/\Gamma^2}$, $B'^2/8\pi$ is the magnetic
energy density. As usual, we assume that a fraction $\epsilon_e$
($\epsilon_B$) of the shock energy density is converted into energy
of relativistic electrons (magnetic field).

To complete the calculations we need the location and the Lorentz
factor of the blast wave as a function of time. The dynamics of the
blast wave is obtained by solving the differential equations
presented by Huang et al. (2000). The possible (but poorly
understood) sideways expansion of the ejecta is ignored. We then
calculate the electron distribution using eq. (\ref{eq:Fan0}) and
the supplemental relations. The quantity $n_{\nu'}$ needed in
eq.(\ref{eq:Jones1}) is calculated via (for simplicity, we consider
only single scattering):
\begin{equation}
n_{\nu'} \approx {T'\over h\nu'}{\sqrt{3}\pi e^3B'\over 4m_e
c^2}\int^{\gamma_{M}}_{\gamma_{\rm min}} {n({\gamma_e})}F({\nu'
\over \nu'_s})d\gamma_e, \label{eq:n_nu}
\end{equation}
where  $n_{\gamma_e} \approx (4\Gamma+3)N_{\gamma_e}/(4\pi R^3/3)
\approx 3\Gamma N_{\gamma_e} /(\pi R^3)$ (The term $4\Gamma+3$ is
introduced by the shock jump condition),
$T'\approx R/(12\Gamma c)$ is the time that the synchrotron
radiation photons stay within the shocked medium, and $\gamma_{\rm
min}\sim 3$ is the Lorentz factor below which the synchrotron
approximation becomes invalid.

Once we know the energy distribution of the electrons, we calculate
the synchrotron and inverse Compton emission, including
synchrotron-self-absorption, and we integrate the observed flux over
the ``equal-arrival time surfaces" (Rees 1966; Waxman 1997; Sari
1998; Granot, Piran \& Sari 1999). In the current code, we did not
take into account the influence of the synchrotron-self-absorption
on the electron distribution, as that done in Pe'er \& Waxman
(2005). However, with typical GRB afterglow parameters adopted in
this work, for $10^2 {\rm sec}<t<10^{5}{\rm sec}$ (at late times,
the high energy emission are usually too low to be of our interest),
it is straightforward to show that the random Lorentz factor of the
electrons emitting at the synchrotron-self-absorption frequency
\cite{CL00,Sari01} is $<100$. The modification of the low energy
electron's distribution through the synchrotron-self-absorption is
thus unlikely to influence the high energy spectrum significantly.

In Fig. 2, we compare the spectral distributions calculated via
the simple instantaneous approach of sec. 3.1 and the more
detailed dynamical approach of this subsection.  The two methods
are clearly consistent with each other. This gives us confidence
in the validity of both calculations. Note that the multiple
inverse Compton scattering are ignored in our dynamical approach
but are included in the Instantaneous approximation. The
consistence between these two approaches suggests that the
multiple inverse Compton scattering is not important, at least for
the typical GRB afterglow parameters (see also Sari \& Esin 2001).

In the case of EIC, the seed photon energy distribution is not
influenced by the electron energy distribution.  From this point of
view the calculations are simpler.  However there is another
complication, viz., for the cases of interest to us, the seed photons
are highly anisotropic in the rest frame of the blast wave. The
spectrum of radiation scattered at an angle $\theta_{\rm sc}$ relative
to the direction of the photon beam penetrating through this region
is \cite{AA81}:
\begin{eqnarray}
{d N_\gamma \over  dt d \nu'_{_{\rm EIC}} d\Omega'} \approx
{3\sigma_T c  \over 16\pi \gamma_e^2}{n_{\nu'} d\nu' \over \nu'} [1+
{\xi^2 \over 2(1-{\xi})}-{2\xi \over b_\theta (1-\xi)}+{2\xi^2 \over
b_\theta^2 (1-\xi)^2}], \label{eq:AA81}
\end{eqnarray}
where $d\Omega'=2\pi \sin \theta_{\rm sc} d\theta_{\rm sc}$, $\xi
\equiv h\nu'_{_{\rm EIC}}/(\gamma_e m_e c^2)$, $b_\theta=2(1-\cos
\theta_{\rm sc})\gamma_e h\nu'/(m_e c^2)$, $\cos \theta_{\rm
sc}=(\cos \theta-\beta)/(1-\beta \cos \theta)$, $\theta$ is the
angle between the line of sight and the emitting point, $\beta$ is
the velocity of the emitting point, and $h\nu'\ll h\nu'_{_{\rm EIC}}
\leq \gamma_e m_e c^2 b_\theta /(1+b_\theta)$.  As expected, on
integration over $\theta_{\rm sc}$, eq. (\ref{eq:AA81}) reduces to
eq. (\ref{eq:Jones1}). The energy loss rate of the hot electron beam
can be estimated by eq. (\ref{eq:P_compt}) and $Y(\gamma_e)$ is
governed by eqs. (\ref{eq:Y_gamma}-\ref{eq:Jones1}) for a given
$n_{\nu'}$ (see \S\ref{sec:EIC} for details).


\section{High energy SSC afterglow}

The dominant source of long lasting high energy GRB afterglow
emission is SSC of the hot electrons in the forward external shock.
At early stages  when the cooling of most electrons is important,
the luminosity of the SSC emission, $L_{_{\rm SSC}}$, is related to
the luminosity of the synchrotron radiation, $L_{_{\rm syn}}$, as
\cite{Sari01}:
\begin{equation}
L_{_{\rm SSC}} \sim {\cal Y} L_{\rm syn},
\end{equation}
where ${\cal Y}$ is the Compton parameter. The X-ray luminosity
$L_{X}$ is a small fraction of $L_{_{\rm syn}}$ but we can use it
as a proxy for the total luminosity. To do so we define a factor
$\epsilon_{\rm X}$ such that $L_{X}\equiv \epsilon_{\rm X} L_{\rm
syn}$ and:
\begin{equation}
L_{_{\rm SSC}} \sim {\cal Y} L_{X}/\epsilon_{\rm X}.
\end{equation}
As long as $\epsilon_{\rm X}$ does not vary significantly with time,
we expect the broad-band SSC afterglow light curve and the X-ray
light curve to have a similar temporal behavior.  We expect,
therefore, that $L_X$ and $L_{\rm SSC}$ should be highly correlated.
This is, of course, confirmed by more detailed analysis, as shown
below in eq. (\ref{eq:Lssc_Lx}).

The light curve depends on the dynamics of the blast wave and in
particular on the evolution with time of $L_{_{\rm eln}}$, the power
given to the shocked electrons (see eq. \ref{eq:L_electron}). We
consider first the evolution expected in the standard afterglow model
and then discuss various modifications to the model.

\subsection{Analytic Considerations}
We begin with the standard afterglow. We consider a circumburst
medium with a number density profile $n_{\rm m}=n_* R^{-k}$,
$0\leq k<3$; here, $k=0$ corresponds to a constant density ISM,
and $k=2$ to a standard stellar wind (M\'esz\'aros, Rees \& Wijers
1998; Dai \& Lu 1998b; Chevalier \& Li 2000), though $k\sim 1.5$
is still possible, as found in some supernovae \cite{Weil02} and
in GRB 991208 \cite{DG01}. The quantity $n_*$ is the number
density at a distance $R=1$:
\begin{equation}
n_* = \left\{%
\begin{array}{ll}
   n,    ~~~~ & \hbox{for k=0,} \\
    3.0\times 10^{35}~ A_* ~{\rm cm^{-3}}, ~~~~& \hbox{for k=2,} \\
\end{array}%
\right. \label{eq:Main1}
\end{equation}
where $A_*=[\dot{M}/10^{-5}M_\odot~{\rm yr^{-1}}][v_w/(10^8{\rm
cm}~{\rm s^{-1}})]$, $\dot{M}$ is the mass loss rate of the
progenitor, $v_w$ is the velocity of the stellar wind \cite{CL00}.

Following the standard afterglow model, we assume that the dynamical
evolution of the ejecta has a Blandford-McKee self-similar profile
\cite{BM76}. The power given to the freshly shocked electrons,
$L_{_{\rm eln}}$, in the blast wave is:
\begin{eqnarray}
L_{_{\rm eln}} \approx ~
\epsilon_{e,-1} E_{k,53} t_3^{-1}\left\{%
\begin{array}{ll}
    7.5\times 10^{48}~{\rm erg~s^{-1}}, & \hbox{for $k=0$;} \\
    ~~5\times 10^{48}~{\rm erg~s^{-1}}, & \hbox{for $k=2$.} \\
\end{array}%
\right. \label{eq:L_electron}
\end{eqnarray}
where $E_k$ is the equivalent isotropic energy of the ejecta.  We
note that $L_{_{\rm eln}}$ depends only weakly on the density
profile. Here and throughout this text, the convention $Q_x=Q/10^x$
has been adopted in cgs units.

The SSC luminosity can be estimated as:
\begin{equation}
L_{_{\rm SSC}}\approx   \epsilon_{_{\rm high}} L_{_{\rm eln}}.
\label{eq:gen5}
\end{equation}
All the physics in this equation is, of course, hidden in the factor
$\epsilon_{_{\rm high}}$, which depends, in turn, on the synchrotron
and cooling frequencies, $\nu_m$ and $\nu_c$ \cite{Sari98}, and on the
power law index of the electron distribution, $p$:
\begin{equation}
\epsilon_{_{\rm high}} \sim \eta {\cal Y}/(1+{\cal Y}),
\end{equation}
where \cite{Sari96,Sari01}:
\begin{eqnarray}
  \eta &\equiv& \min \{ 1, (\nu_m/ {\nu}_c)^{(p-2)/2} \}, \\
  {\cal Y} &\sim & (-1+\sqrt{1+4 \bar{\eta}
\epsilon_e/\epsilon_B})/2, \\
  \bar{\eta} &\equiv & \min \{ 1, (\nu_m/
\bar{\nu}_c)^{(p-2)/2} \}, \\
  \bar{\nu}_c &=& (1+{\cal Y})^2 \nu_c \ .
\end{eqnarray}
The ratio of the synchrotron and  cooling frequencies satisfies
\cite{Sari98,Yost03}:
\begin{equation}
{\nu_m \over \bar{\nu}_c} \approx \left\{%
\begin{array}{ll}
    0.0024
C_p^2 \epsilon_{e,-1}^2 \epsilon_{B,-2}^2 n E_{k,52} t_3^{-1}, & \hbox{for $k=0$,} \\
    0.12
C_p^2 \epsilon_{e,-1}^2
\epsilon_{B,-2}^2 A_{*,-1}^2 t_3^{-2}, & \hbox{for $k=2$,} \\
\end{array}%
\right.
\end{equation}
where $C_p \equiv 13(p-2)/[3(p-1)]$.  Note that, in all analytical
relations, the time and the frequency are measured in the burst's
frame, that is we ignore cosmological (1+z) corrections. Numerical
results are presented for a canonical burst at $z=1$.

The X-ray band is typically above  $\max \{ \nu_m, \nu_c\}$. In this
case the forward shock X-ray emission can be related to the kinetic
energy of the forward shock \cite{K00,FW01,FP06a}:
\begin{eqnarray}
L_X  \approx
\epsilon_{\rm B,-2}^{\rm (p-2)/4} \epsilon_{\rm e,-1}^{\rm p-1}
(1+{\cal Y})^{-1}E_{k,53}^{\rm (p+2)/4}
{t_3}^{(2-3p)/4} \left\{%
\begin{array}{ll}
    8.8\times 10^{47}~{\rm ergs~s^{-1}}, & \hbox{for $k=0$;} \\
    1.4\times 10^{48}~{\rm ergs~s^{-1}}, & \hbox{for $k=2$.} \\
\end{array}%
\right. \label{eq:Lx}
\end{eqnarray}
We thus have
\begin{eqnarray}
{L_{_{\rm SSC}}\over L_X} \sim  4 \eta {\cal
Y}\epsilon_{e,-1}^{2-p}\epsilon_{B,-2}^{(2-p)/4}
 E_{k,53}^{(2-p)/4} t_3^{3(p-2)/4} \left\{%
\begin{array}{ll}
    2, & \hbox{for $k=0$;} \\
    1, & \hbox{for $k=2$,} \\
\end{array}%
\right. \label{eq:Lssc_Lx}
\end{eqnarray}

For a universal $p\sim 2.1-2.3$, $L_{_{\rm SSC}}/L_X$ is sensitive
only to $\eta$ and ${\cal Y}$ and it is only weakly dependent on other
parameters. At early times, when the cooling of electrons is
important, $L_{_{\rm SSC}} \propto t_3^{3(p-2)/4}L_X$ (note that in
the standard afterglow model, $E_k,~n_*,~\epsilon_e,~\epsilon_B$ are
all constant). Therefore a wide band SSC light curve will have a
temporal behavior quite similar to that of the X-rays.

\begin{figure}
\begin{picture}(0,200)
\put(0,0){\includegraphics{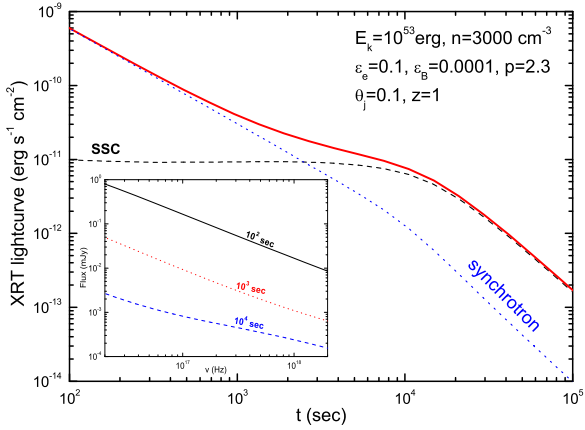}}
\end{picture}
\caption{The XRT light curve in a dense ISM. The thick solid line
is the predicted X-ray light curve, including both the synchrotron
and the SSC components of the forward shock. The insert shows the
corresponding X-ray spectra at three different times, as marked in
the plot.} \label{fig:XRT}
\end{figure}

Roughly speaking, the energy of the SSC emission peaks at a
frequency $\sim \max \{\nu_m^{\rm SSC}, \nu_c^{\rm SSC}\}$, where
$\nu_m^{\rm SSC} \approx 2\gamma_m^2 \nu_m$ and $\nu_c^{\rm SSC}
\approx 2\gamma_c^2 \nu_c$, where $\gamma_{c}$ is the cooling
Lorentz factor of shocked electrons. Following the standard
treatment \cite{Sari98,CL00}, we have
\begin{eqnarray}
\nu_m^{\rm SSC} \approx  10^{21}~{\rm Hz}~ C_p^4 \epsilon_{e,-1}^4
\epsilon_{B,-2}^{1/2}
\left\{%
\begin{array}{ll}
    6.2~
 n^{-1/4} E_{k,53}^{3/4} t_3^{-9/4}, & \hbox{for $k=0$;} \\
    1.4~
 A_{*,-1}^{-1/2} E_{k,53} t_3^{-2}, & \hbox{for $k=2$,} \\
\end{array}%
\right.
\end{eqnarray}

\begin{eqnarray}
\nu_c^{\rm SSC} \approx  10^{24}~{\rm Hz}~ (1+{\cal Y})^{-4}
\epsilon_{B,-2}^{-7/2}
 \left\{%
\begin{array}{ll}
    4~
 n^{-9/4} E_{k,53}^{-5/4} t_3^{-1/4}, & \hbox{for $k=0$;} \\
    1.5~
 A_{*,-1}^{-9/2} E_{k,53} t_3^{2}, & \hbox{for $k=2$.} \\
\end{array}%
\right.
\end{eqnarray}
Note that $\nu_c^{\rm SSC} \propto n^{-9/4}$ or $\propto
A_*^{-9/2}$. So $n\sim 10^3~{\rm cm^{-3}}$ or $A_* \sim {\rm a
~few}$ will shift $\nu_c^{\rm SSC}$ to the X-ray/UV/optical band and
in this case the SSC emission will influence the X-ray observations.
It may even cause a flattening of the X-ray light curve due to the
emergence of this new component.
 An example of such a case is shown in Fig.\ref{fig:XRT}. One can
see a flat X-ray segment, which is rather similar to that detected
by {\it Swift}. However, it is not clear that this  can account for
the {\it Swift} observations because in this case  the X-Ray
spectrum would vary with time (see the insert of Fig.\ref{fig:XRT}).
Such variations are not seen in the {\it Swift} data.

The (adiabatic) standard afterglow model assumes that (i) the
outflow energy is a constant and (ii) the shock parameters are
constant. As mentioned earlier this model is inconsistent with the
shallow decline phase (phase-II). One possibility is that one of
these two basic assumptions should be revised
\cite{Zhang06,FP06a,Ioka06,Panaitescu06,Nousek06,GKP06}. We consider
energy injection of the form $E_{k} \propto t^{1-q}$
\cite{Cohen99,ZM01a}, where $q=1$ represents no energy injection and
$q=0$ corresponds to a pulsar/magnetar-like energy injection (Dai \&
Lu 1998; Zhang \& M\'esz\'aros 2001a; Dai 2004; Fan \& Xu 2006).
Other $q$ values are possible for an energy injection that results
from slower material progressively catching up (Rees \& M\'esz\'aros
1998; Kumar \& Piran 2000; Sari \& M\'esz\'aros 2000; Granot \&
Kumar 2006) or if an energy injection is caused by the fall-back of
the envelope of the massive star (MacFadyen, Woosley \& Heger 2001;
Zhang, Woosley \& Heger 2007). We also explore the situation where
the equipartition parameters, $\epsilon_e$ and $\epsilon_B$, are
shock-strength dependent (i.e., time dependent)\footnote{One may
speculate that the energy distribution index of the accelerated
electrons $p$ is also time-evolving. However, this is not seen in
the data as the spectrum does not vary during this phase.}, though
the underlying physics is far from clear \cite{PF06}. Instead of
exploring the possible physical processes that lead to such a
phenomenon, we simply take $(\epsilon_e,~\epsilon_B) \propto
(t^{c},~t^{d})$. These modifications lead to:
\begin{equation}
L_X \propto {\eta \epsilon_{\rm X} \over 1+{\cal Y} } t^{(c-q)}.
\end{equation}

It is straightforward to show that
\begin{equation}
L_{_{\rm SSC}} \propto {\eta {\cal Y} \over 1+{\cal Y} } t^{(c-q)},
\label{eq:gen3a}
\end{equation}
\begin{equation}
\nu_m^{_{\rm SSC}} \propto t^{4c+{d \over
2}+{(6-k)(1-q)+(5k-18)\over 2(4-k)}},
\end{equation}
\begin{equation}
\nu_c^{_{\rm SSC}} \propto t^{-{7\over 2}d+{6(k-1) \over
4-k}-{(7k-10)\over 2(4-k)}q}(1+{\cal Y})^{-4}.
\end{equation}
Equation (\ref{eq:gen3a}) is one of our main results. As expected,
with significant energy injection, or either $\epsilon_e$
increasing with time, or both, $L_{_{\rm SSC}}{\rm (general)}$ is
flattened.  An $\epsilon_B$ decreasing (increasing) with time will
also flatten (steepen) the high energy emission light curve.
However, such a modification seems to be small and it cannot give
rise to either the observed shallow decline phase of the x-ray
light curve or to a detectable signature in the high energy
component. Therefore, we focus on models with either
time-dependent $E_k$ or $\epsilon_e$.

The shallow decline seen in the X-ray light curve during phase-II
(Fig. 1) requires $q\sim 0.5$ or $c \sim 0.4$.  In general, for $L_X
\propto t^{-\alpha}~(\alpha \leq 1)$ we need (in the energy injection
case) $q = [4(\alpha+1)-2p]/(p+2)$ which yields $L_{_{\rm SSC}}
\propto t^{-[4(\alpha+1)-2p]/(p+2)}\propto t^{-\alpha}$ for $p\sim
2$. The high energy decline is quite similar to the decline of the
X-rays.  For a varying $\epsilon_e$ (with no energy injection, i.e.,
$q=1$), we need $c=(3p-2-4\alpha)/[4(p-1)]$, which in turn results in
$L_{\rm SSC} \propto t^{(p+2-12\alpha)/[8(p-1)]}\propto
t^{(1-3\alpha)/2} ~{\rm for}~p\sim 2$. The high energy decline is
slightly slower than that of the X-rays in this case.

At least in principle, one could combine IR/optical/UV/X-ray and
high energy observations to distinguish between the two
modifications described above.  For example, we have $\nu_m
\propto \epsilon_e^2 E_k^{1/2}$ and $\nu_m^{_{\rm SSC}} \propto
\epsilon_e^4 E_k^{(6-k)/[2(4-k)]}$. If the early X-ray flattening
was caused by $\epsilon_e \propto t^c$, $\nu_m$ and $\nu_m^{_{\rm
SSC}}$ will decline much more slowly than in the energy injection
case $E_k \propto t^{1-q}$. The wide energy range of LAT onboard
GLAST (20MeV-300GeV) might enable us to observe the variations of
$\nu_m^{\rm SSC}$ with time.

It is interesting to note in passing that these modifications
provide a possible explanation for some long-term puzzles in GRB
940217. The long-lasting MeV to GeV afterglow emission of GRB 940217
\cite{Hurley94} showed two remarkable features: (a) the count rate
of high energy photons was almost a constant (b) the typical energy
of these photons was nearly unchanged. These two features can be
reproduced with $c \sim q \sim 1/2$ and $d=k=0$ (Wei \& Fan 2007).


\subsection{Numerical Results}

\begin{figure}
\begin{picture}(0,300)
\put(0,0){\includegraphics{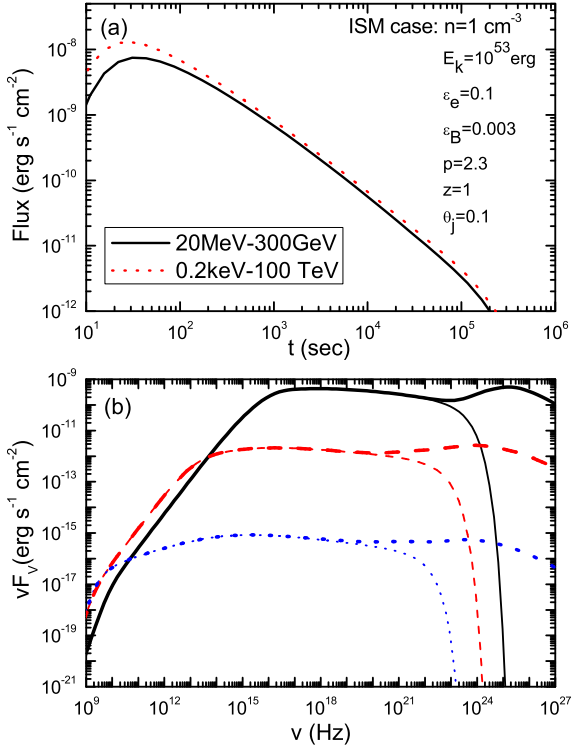}}
\end{picture}
\caption{SSC radiation from the forward shock for the case of a
constant density external ISM. (a) SSC light curves in the energy
ranges 20MeV-300GeV and 0.2 keV -100 TeV, respectively. (b) Spectra at
three selected times; thin and thick lines correspond to the pure
synchrotron spectrum and SSC+synchrotron spectrum, respectively, and
solid, dashed and dotted lines are at $2 \times 10^2$, $2 \times 10^4$
and $2 \times 10^6$ s after the burst. }
\label{fig:SSC1a}
\end{figure}

\begin{figure}
\begin{picture}(0,300)
\put(0,0){\includegraphics{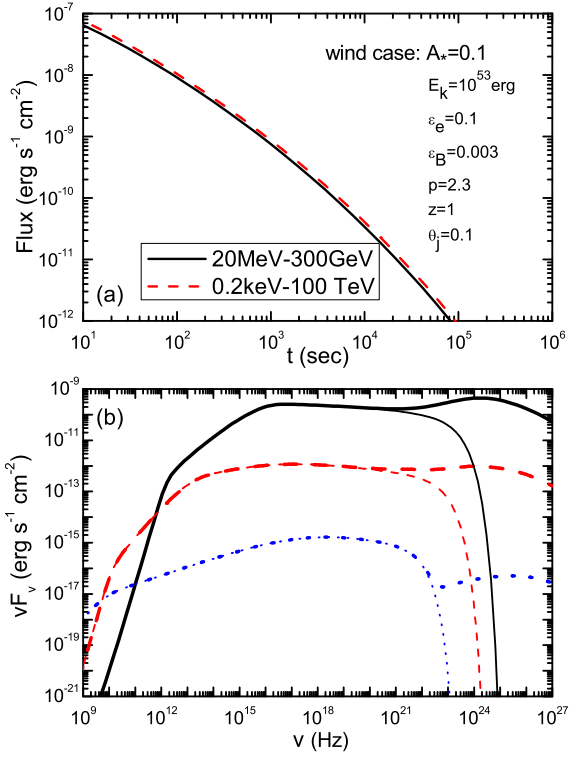}}
\end{picture}
\caption{SSC radiation from the forward shock for the case when the
external medium corresponds to the wind from the progenitor star
($k=2$). The line styles are the same as in Fig. \ref{fig:SSC1a}.}
\label{fig:SSC1b}
\end{figure}

We turn now to numerical computations of the high energy light
curves. We consider, first, the standard afterglow model using typical
parameters that seem to fit the average late afterglow: $E_k =
10^{53}$ erg, $p=2.3$, $\epsilon_e=0.1$, $\epsilon_B = 0.003$ and
$\theta_j = 0.1$. We consider a typical burst at $z = 1$. Figures
\ref{fig:SSC1a} and \ref{fig:SSC1b} depict the calculated light curves
and spectra for two models of the external medium: a uniform density
ISM and a stellar wind. In both figures, panel (a) shows the SSC
emission afterglow light curve and panel (b) shows the spectrum.

\begin{figure}
\begin{picture}(0,170)
\put(0,0){\includegraphics{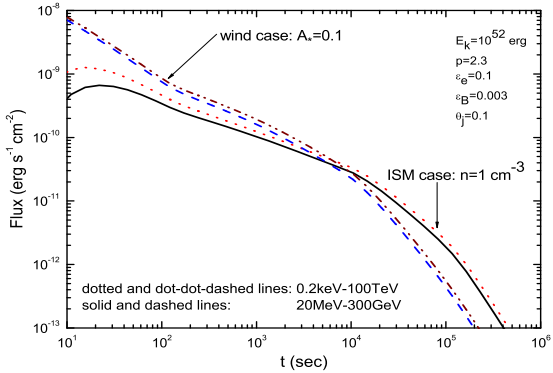}}
\end{picture}
\caption{SSC radiation from the forward shock for the case when energy
from the central engine is injected over a period of time: $dE_{\rm
inj}/dt =5\times 10^{49}(t/100{\rm s})^{-0.5}~{\rm erg~s^{-1}}$ for
$10^2~{\rm s}<t<10^4~{\rm s}$. Note that the SSC emission light curve
flattens as a result of the energy injection.} \label{fig:SSC_inj}
\end{figure}

We consider now an energy injection model where the energy injection
has the form:
\begin{equation}
dE_{\rm inj}/dt =5\times 10^{49}(t/100{\rm s})^{-0.5}~{\rm
erg~s^{-1}}
\end{equation}
for $10^2~{\rm s}<t<10^4~{\rm s}$, which corresponds to $q=0.5$ and
$E_k = 10^{52}$ erg. Apart from $q$ and $E_k$ all other parameters are
similar to those used in the standard case above. The total integrated
energy injected is equal to $9\times 10^{52}{\rm erg}\gg E_k$.  The
resulting light curve is shown in Fig. \ref{fig:SSC_inj}. The SSC
light curve is flattened when the energy injection is strong enough to
suppress the deceleration of the outflow.  The numerical light curve
has $L_{\rm SSC}\propto t^{-(0.6\sim 0.7)}$ which is consistent with
our analytic estimate $L_{\rm SSC}\propto t^{-0.5}$ for $c=0$ and
$q=0.5$ (see eq. (\ref{eq:gen3a})).

\begin{figure}
\begin{picture}(0,170)
\put(0,0){\includegraphics{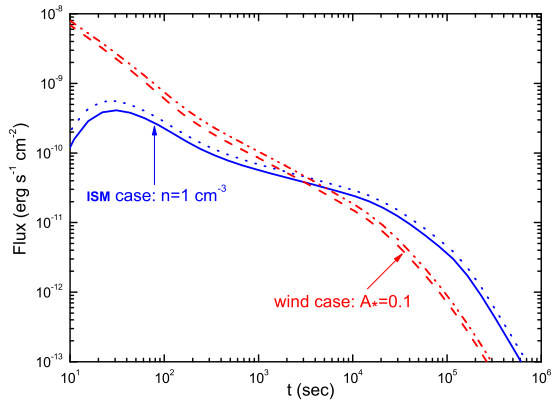}}
\end{picture}
\caption{SSC radiation from the forward shock for the case when the
electron energy parameter $\epsilon_e$ varies with time. The solid and
dashed lines correspond to the emission in the energy range
20MeV-300GeV, while the dotted and dot-dot-dashed lines are for the
emission in the energy range 0.2 keV-100 TeV. The shock parameters are
$\epsilon_B=0.003$, $\epsilon_e=0.017$ for $t<100$ s,
$\epsilon_e=0.017(t/10)^{0.4}$ for $t<10^4$ s after which it
saturates.  Other parameters are $E_k=10^{53}$ erg, $z=1$,
$\theta_j=0.1$, $p=2.3$. The parameters corresponding to the external
mdium are marked on the plot.} \label{fig:SSC_evo}
\end{figure}

We turn now to a time-evolving shock parameter $\epsilon_e$, and
consider $\epsilon_e$ varying as $t^{0.4}$. As shown in eq.
(\ref{eq:gen3a}) and in Fig. \ref{fig:SSC_evo}, an increase with time
of $\epsilon_e$ flattens the high energy emission light curve. The
very small $\epsilon_e$ at early time not only lowers the fraction of
the shock energy given to the fresh electrons but it also suppresses
the SSC emission.  The resulting $t^{-0.5}$ decline depicted in
Fig. \ref{fig:SSC_evo} is consistent with the analytic estimate
$t^{-0.4}$ for $c=0.4$ (and $q=1$).

\begin{figure}
\begin{picture}(0,170)
\put(0,0){\includegraphics{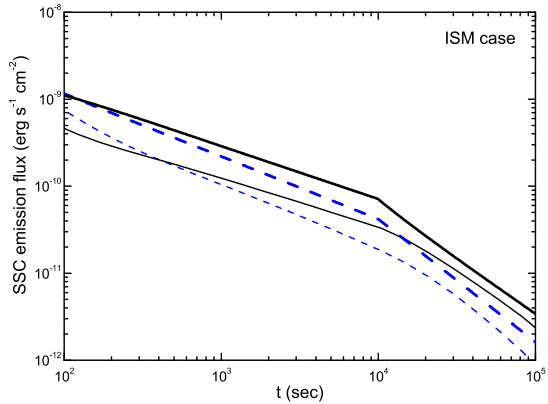}}
\end{picture}
\caption{Comparison of numerical and analytical results for ISM. The
thin lines are our numerical SSC light curves for the energy range
$0.2{\rm keV}-100{\rm TeV}$, while the thick lines are the
corresponding analytical results. The solid and dashed lines are the
results in the cases of energy injection and $\epsilon_e$ increasing
with time, respectively. The parameters are the same as in Fig.
\ref{fig:SSC_inj} and in Fig. \ref{fig:SSC_evo}, respectively. For
the wind medium, the results are rather similar.} \label{fig:check}
\end{figure}

To check the consistency of the numerical and analytic results, we
plot the two estimates of $L_{_{\rm SSC}}$ (using eq. \ref{eq:gen5})
in Fig. \ref{fig:check}. The analytic results (the thick lines) are a
factor of $2-4$ times larger than the corresponding numerical results
(the thin lines).  This is reasonable as some important corrections,
such as the integration of the emission over ``equal-arrival
surfaces", have been ignored in the analytic formulae.

\section{High energy emission associated with X-ray flares}

We turn now to GeV flares that might arise from inverse Compton
scattering of the radiation associated with X-ray (or UV) flares.
Although X-ray flares (phase-V in Fig.\ref{fig:Cartoon}) were detected
even before {\it Swift}, their frequency became clear only after {\it
Swift} began its observations. By now it is known that flares are
quite common and can appear at all phases of the afterglow. At times
the energy emitted in a flare can be fairly large. There are two main
ideas to explain the origin of these flares: (i) ``Late internal
shocks" (Fan \& Wei 2005, Burrow et al. 2005, Zhang \etal 2006)
associated with a long-lived central engine, and (ii) ''refreshed
shocks" \cite{Guetta07} when late shells encounter the external shock
and lead to brightening.

Inverse Compton scattering of photons from an X-ray flare are possible
via two distinct mechanisms.  It could be the result of SSC emission
from the same electrons that produce the X-ray flares.  If the X-ray
flare is produced by late internal shocks, then an additional source
of high energy radiation is possible, viz., EIC scattering of flare
photons by hot electrons in the external shock.  We consider both
possibilities.

\subsection{SSC flares}
\label{sec:GeV_flash}

SSC within the same shock that produces the X-ray flare will give a
high energy flare simultaneously with the low energy flare. This would
arise if the X-ray flare results from either a late internal shock or
from a refreshed shock within the forward external shock.

It is difficult to predict the expected SSC emission as we have no
robust estimate of the typical Lorentz factor of the shocked
electrons that produce the X-ray flare. A critical factor is the
location of the shock, which determines the various parameter
within the emitting region. For prompt $\gamma-$rays from an
internal shock, the typical radius of the shock is $R_{\rm
prompt}\sim 10^{13}-10^{15}$cm \cite{Piran99,Piran04}.  If flares
are produced by ``late internal shocks", $R_{\rm flare}\sim
10^{15}$ cm is possible \cite{FW05}, whereas with ``refreshed
external shocks", $R_{\rm flare}$ may be as large as $10^{17}$ cm
\cite{Galli06b,Wu06,Guetta07}.

Assuming  that the soft X-ray flares are powered by the synchrotron
radiation of the shocked electrons we can estimate the typical
Lorentz factor of the electrons, $\gamma_{\rm e,m}$. The magnetic
field, $B$, at $R_{\rm flare}$ can be estimated by:
\begin{eqnarray}
 B \sim  [2\varepsilon L_{X}/(\Gamma^2 R_{\rm flare}^2 c)]^{1/2}
\sim  250~{\rm Gauss}~\varepsilon^{1/2} L_{\rm
x,49}^{1/2}\Gamma^{-1}R_{\rm flare,17}^{-1},
\end{eqnarray}
where $\varepsilon \equiv \epsilon_B/\epsilon_e$. For this value of
the magnetic field, the peak energy of the flare photons will be at
$E_{\rm p} \sim 0.2$ keV if the typical electron Lorentz factor is
\begin{equation}
\gamma_{\rm e,m} \sim 800~\varepsilon^{-1/4}L_{X,49}^{-1/4} R_{\rm
flare,15}^{1/2}(E_{\rm p}/0.2~{\rm keV})^{1/2}.
\label{eq:gamma_em}
\end{equation}
The energy of a typical inverse Compton photon is then
\begin{eqnarray}
h\nu_p^{\rm ssc} \sim  2\gamma_{\rm e,m}^2 h\nu_p \sim 0.3 {\rm
GeV}~\varepsilon^{-1/2}L_{X,49}^{-1/2} R_{\rm flare,15}(E_{\rm
p}/0.2~{\rm keV})^{2}. \label{eq:SSC_flare}
\end{eqnarray}
Thus, high energy emission simultaneous with the X-ray flare is
expected if the emitting region is not significantly magnetized.

Roughly speaking, the total fluence of the SSC emission of the flare
shock is comparable to that of the X-ray emission, typically
$10^{-7}\sim 10^{-6}~{\rm erg~cm^{-2}}$ integrated over the
aftgerglow. In late internal shocks (i.e., $R_{\rm flare}\sim
10^{15}$ cm), a GeV flash accompanying the X-ray flare is possible
(Wei, Yan \& Fan 2006). This problem was also discussed by Wang et
al. (2006), who assumed that $\gamma_{\rm e,m}\sim 100$ and obtained
$h\nu_p^{\rm ssc} \sim 10$ MeV, which they considered as
uninteresting. However, as shown above, $\gamma_{\rm e,m}$ can be
large to $\sim 1000$ and $h\nu_p^{\rm ssc}$ is two orders of
magnitude larger. In refreshed external shocks, a GeV-TeV flash is
predicted because of its very large $R_{\rm flare}~(\sim 10^{17}$
cm), as indicated in eq. (\ref{eq:SSC_flare}). More detailed
analysis can be found in Galli \& Piro (2007).

A subtle issue that has to be checked is whether the high energy
photons will be absorbed by pair production on the high energy tail of
the flare.  The pair production optical depth for photons with energy
$E_{\rm cut}$ (absorbed by the flare photons with energy $E_{a}\sim
2(\Gamma m_e c^2)^2/E_{\rm cut}\sim 0.5 {\rm
MeV}~\Gamma_{1.5}^{2}(E_{\rm cut}/1{\rm GeV})^{-1}$) can be estimated
as (e.g., Svensson 1987)
\begin{eqnarray}
\tau_{\gamma \gamma}  \simeq  {11 \sigma_T N_{>E_{a}} \over 720 \pi
R_{\rm flare}^2} \sim  4\times 10^{-2}~R_{\rm flare,15}^{-2} F_{\rm
flare,-8.3} \delta t_{1} D_{L,28.34}^2 ({E_{\rm p}\over 0.2 {\rm
keV} })^{\beta_{\rm flare}-1}\Gamma_{1.5}^{-2\beta_{\rm
flare}}({E_{\rm cut}\over 1{\rm GeV}})^{\beta_{\rm flare}},
\label{eq:Optdepth}
\end{eqnarray}
where $N_{>E_a}= {\beta_{\rm flare}-1 \over \beta_{\rm flare}} ({
E_{\rm p} \over E_{a} })^{\beta_{\rm flare}}{4\pi D_L^2{F_{\rm
flare} \delta t}\over E_{\rm p}}$ is the total flare photon number
of one pulse satisfying $h \nu>E_a$, where $\delta t$ is the
timescale of the flare pulse and the high-energy power-law index
$\beta_{\rm flare}\sim 1.2$ has been used to get the numerical
coefficient. Clearly, for $R_{\rm flare}\sim 10^{17}$cm, i.e., the
refreshed shock case, the tens GeV high energy photon emission will
not be absorbed by the flare photons. For $R_{\rm flare} \sim
10^{15}$ cm, i.e., the late internal shock case, the small optical
depth will not affect the sub-GeV flux unless $\delta t_1 >25
\Gamma_{1.5}^{2\beta_{\rm flare}}R_{\rm flare,15}^2$.

\subsection{Extended EIC plateau}\label{sec:EIC}

We turn now to the scenario in which the X-ray flares are produced
by late internal shocks (Fan \& Wei, 2005, Burrow et al. 2005, Zhang
\etal 2006). We calculate the inverse Compton scattering of these
seed photons by hot electrons accelerated within the external shock.
We assume that the X-ray flares are accompanied by far-UV emission
and calculate the upscattering of these photons as well. A central
ingredient of this scenario is that in the rest frame of the blast
wave, the seed photons are highly beamed. We take care of this
effect, following the analysis of Aharonian \& Atoyan (1981).

If the EIC emission is simultaneous with the X-ray flare (i.e.,
the duration of the EIC emission has not been extended
significantly), the EIC luminosity can be estimated by eq.
(\ref{eq:gen5}). However, in the rest frame of the shocked
material, the EIC emission has a maximum at $\theta_{\rm sc}=\pi$
and it vanishes for small scattering angles \cite{AA81,Brun01}.
This effect lowers the high energy flux in two ways. First, a
fraction of the total energy is emitted out of our line of sight
and thus the received power is depressed (relative to the
isotropic seed photon case). This yields a correction by a factor
of 2 (which we ignore henceforth).  Second and more important, the
strongest emission is from\footnote{This could be more easily
understood in the Thomson regime. As shown in eq.(43) of Brunetti
(2001), in the local frame of the shocked medium, the emissivity
is proportional to $(1-\cos \theta_{\rm sc})^{(1+\delta)/2}$,
where $\delta=p~{\rm or}~p+1$, depending on the cooling of the
electrons. The observed emission from an angle $\theta$ is thus
$\propto [\Gamma (1-\beta \cos \theta)]^{-3} \sin \theta (1-\cos
\theta_{\rm sc})^{(1+\delta)/2} \propto
\theta^{\delta+2}(1+\Gamma^2 \theta^2)^{-(7+\delta)/2}\equiv {\cal
F}$ since $\cos \theta_{\rm sc}=(\cos \theta -\beta)/(1-\beta \cos
\theta)$. The requirement that $d{\cal F}/d\theta=0$ yields
$\theta \approx \sqrt{(2+\delta)/5}/\Gamma \sim 1/\Gamma$ for
$\delta \sim 2-3$.} $\theta \sim 1/\Gamma$. Thus the peak time of
the high energy EIC emission is estimated to be (Fan \& Piran
2006b; Wang \& M\'esz\'aros 2006):
\begin{equation}
T_{\rm p} \sim
R/(2\Gamma^2c)\sim (4-k)t_{\rm f},  \label{eq:t_p}
\end{equation}
where $t_{\rm f}$ is the time when the X-ray flare ceases. $T_{\rm
p}$, which is also proportional to the duration of the high energy
peak, could be much longer than $\Delta T$, the duration of the soft
X-ray flare.

The luminosity of the  high energy flare would be lower than the
simple estimate by the ratio of the durations:
\begin{equation}
L_{_{\rm EIC}}\sim {L_{\rm eln}\over (T_{\rm p}/\Delta T)}.
\end{equation}
At 100-1000 s after the burst, the forward shock emission peaks in
the far-UV to soft X-ray band, and the corresponding SSC emission
peaks in sub-GeV to GeV energy range. A comparison of the SSC
luminosity of the forward shock after but around $t_{\rm f}$,
$L_{_{\rm SSC}}$ (eq. (\ref{eq:gen5})), with $L_{_{\rm EIC}}$
shows that the SSC emission would be stronger than the EIC
emission and the wide EIC flare would be undetectable.

However, if the forward shock electrons are in the slow cooling regime
before the X-ray flare, their SSC emission is weak and the EIC flare
might be detectable. In this case the total energy available for
extraction in the EIC process $\sim L_{\rm eln} \Delta T + N_e \Gamma
\min \{ \gamma_c, \gamma_m\} m_e c^2$ is much larger than $\sim L_{\rm
eln} \Delta T$, where $N_e$ is the total number of electrons swept by
the forward shock at the time $\sim t_f-\Delta T$ and at the same time
$L_{_{\rm SSC}}$ is much smaller than $L_{\rm eln}$. Though $L_{_{\rm
SSC}}$ may still outshine $L_{_{\rm EIC}}$ at $t \sim t_f$, since it
decreases rapidly with time (steeper than $t^{-1}$, as both $\eta$ and
${\cal Y}/(1+{\cal Y})$ are decreasing with time, see
eq. (\ref{eq:gen5})), the EIC high energy emission may still dominate
at late times.

If the EIC emission dominates over the SSC emission, the high energy
light curve will flatten, as we show below (e.g., Fig.
\ref{fig:EIC_flux}). Such a flattening could arise also as a result of
energy injection or due to an increasing $\epsilon_e$. However, as we
argued in the last section, in those two scenarios, the X-ray and the
high energy emission light curves are quite similar and flattening
should be apparent also in the X-ray signal. The EIC emission should,
on the other hand, show an X-ray flare preceding high energy emission
and not accompanying a flat X-ray light curve.

As an example we consider the  giant flare of GRB 050502b
\cite{Burrows05,Falcone06} and examine the expected external IC
emission that will arise from such a flare. The flux of the flare,
in the 0.2$-$10 keV energy band, can be approximated as a steep
rise: $F_{\rm flare}\approx 5\times 10^{-9}~{\rm
erg~s^{-1}~cm^{-2}}~(t/680{\rm s})^{7}$ for $300{\rm s}<t<680$s, a
constant plateau lasting until $\sim 800$s and a subsequent sharp
decline which might be due to a curvature emission component
\cite{Fenimore96,KP00,Liang06}.  To calculate the EIC emission we
need (see eq. \ref{eq:Jones1}) $n_\nu'$, the distribution of the
seed photons in the rest frame of the shocked medium.

If the flare originates from activity of the central engine
\cite{FW05,Zhang06} one might expect that the radiation process is
similar to that of the prompt emission. Lacking exact information on
the spectrum of the flare and in particular on its peak energy we
assume that it has a typical Band function \cite{Band93}:
\begin{equation} n_{\nu'} =A
\left\{%
\begin{array}{ll}
    ({h \nu' \over 1~ {\rm keV}})^{-(1+\alpha_{\rm flare})}\exp (-\nu'/\nu'_{\rm p}),
&   \hbox{for $\nu' \leq B\nu'_{\rm p}$;} \\
     ({Bh {\nu'}_{\rm p} \over 1~ {\rm keV}})^{\beta_{\rm flare}-\alpha_{\rm flare}}
     \exp(-\beta_{\rm flare}-\alpha_{\rm flare}) ({h\nu' \over 1~ {\rm keV}})^{-(1+\beta_{\rm flare})},
& \hbox{for $B\nu'_{\rm p}\leq \nu'$;} \\
\end{array}%
\right. \label{eq:Band}
\end{equation}
where the high-energy power-law index $\beta_{\rm flare}\approx {\rm
const} \sim 1.2$ and the low-energy power-law index $\alpha_{\rm
flare}\approx {\rm constant} \sim 0$. As the peak energy is not
known we consider three representative values: $E_{\rm p}= 0.02,
0.2, 2$keV.

For a given $\nu'_p$, the parameter $A$ in eq. (\ref{eq:Band}) is
obtained from the observed flux:
\begin{equation}
\int^{\rm 100 keV /h \Gamma}_{\rm 0.02 keV/h \Gamma} n_{\nu'} h\nu'
d\nu'\approx {D_L^2 F_{\rm flare} \over 2R^2 \Gamma^2 c},
\label{eq:Observation}
\end{equation}
where we used lower and upper limits on $h \nu'$ of ${20 {\rm
eV}/\Gamma}$ and $100 {\rm keV}/\Gamma$.  This is because the
self-absorption frequency is likely to be in the UV band \cite{FW05}
and the emission in the hard X-ray band is unknown for nearly all
flares.

\begin{figure}
\begin{picture}(0,260)
\put(0,0){\includegraphics{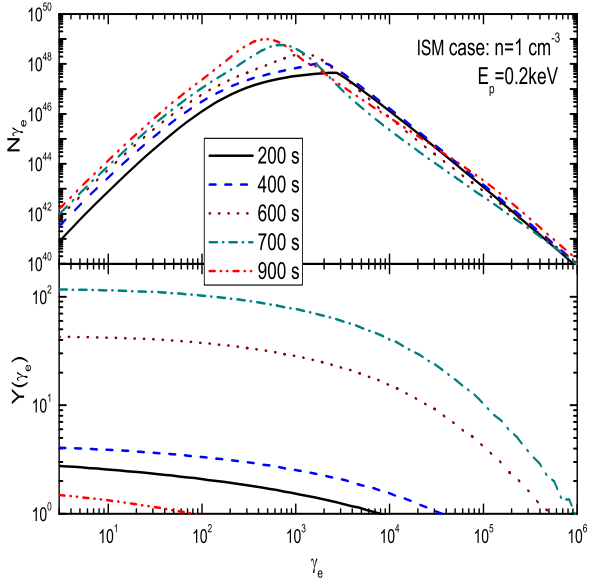}}
\end{picture}
\caption{The electron distribution $N_{\gamma_{\rm e}}$ and the
Compton parameter $Y(\gamma_e)$ as functions of the electron Lorentz
factor $\gamma_{\rm e}$.  The flare photons are assumed to scatter off
electrons in the forward shock for the case of a uniform external
ISM. Different lines represent different times after the initial
burst, as indicated in the figure; the times are determined by
$dt=(1+z)dR/(2\Gamma^2 c)$. The parameters of the model are: $E_{\rm
k} = 10^{52}$ erg, $n = 1~{\rm cm^{-3}}$, $z=1$, $p=2.3$, $\epsilon_e=
0.1$ and $\epsilon_B = 0.01$. The parameters of the flare are
described in the text and the peak energy of the flare emission is
taken to be $E_{\rm p}\sim 0.2$ keV.} \label{fig:EIC_dis}
\end{figure}
\begin{figure}
\begin{picture}(0,260)
\put(0,0){\includegraphics{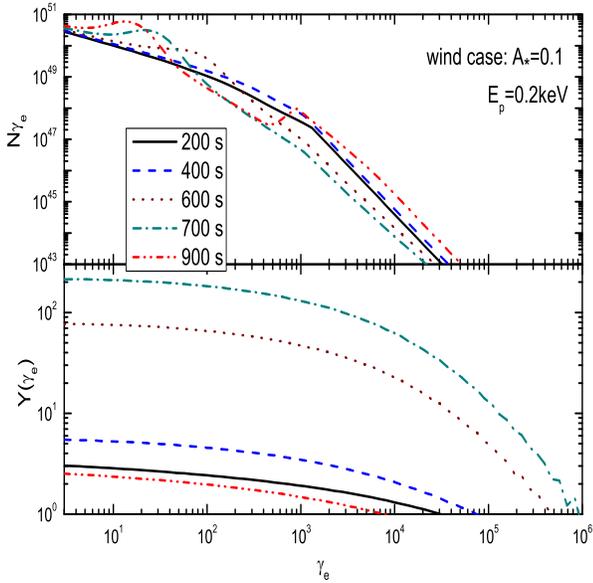}}
\end{picture}
\caption{The electron distribution $N_{\gamma_{\rm e}}$ and the
Compton parameter $Y(\gamma_e)$ as functions of the electron Lorentz
factor $\gamma_{\rm e}$.  Here the flare photons are assumed to
scatter off electrons in the forward shock for the case when the
external medium is due to a stellar wind with $A_*=0.1$.  Other
parameters are the same as in Fig. \ref{fig:EIC_dis}.}
\label{fig:EIC_disb}
\end{figure}

The redshift of GRB 050502b is unknown. We assume the canonical
value of $z=1$ for which $E_{\rm k}\sim 10^{52}$ erg because the
$\gamma-$ray fluence is $\sim 10^{-6}~{\rm erg~cm^{-2}}$
\cite{Burrows05}. For the other parameters we take $n = 1~{\rm
cm}^{-3}$, $p= 2.3$, $\epsilon_e= 0.1$, $\epsilon_B = 0.01$, and
$\theta_j = 0.1$. Figures \ref{fig:EIC_dis} and \ref{fig:EIC_disb}
depict the electron distributions and the Compton parameters as
functions of $\gamma_e$, for $E_{\rm p}\sim 0.2$ keV. The cooling
effect of the X-ray flare photons on the blast wave electrons is
seen clearly in these figures. One sees that the energy of the
electrons is depressed between 400 and 700 s and then it increases
at 900 s when the cooling effect due to the flare photons ceases. As
expected the higher the flare luminosity, the stronger the EIC
cooling. Electrons with $\gamma_e<10^6$ lose most of their energy
via the EIC process (see the large values of the Compton parameter
for these electrons in Figs. \ref{fig:EIC_dis}(b) and
\ref{fig:EIC_disb}(b)).

The resulting high energy emission is shown in Fig.
\ref{fig:EIC_flux}. The SSC emission decreases during and after
the flare as the electrons are cooled by the EIC process. Also at
a later time we get contributions to the observed spectrum from
higher latitude regions from which the emission is weaker.  The
EIC emission is not simultaneous with the X-ray flare. It peaks at
$\sim (4-k) t_f$ (see eq. (\ref{eq:t_p})), and it lasts much
longer than the X-ray flare. This temporal behavior is determined
by the geometry of the emitting surface, the radiation spectrum
and the highly anisotropic EIC emission.  The lagging behavior is
unique and if it is observed, i.e., if it is not hidden by SSC
emission, it would demonstrate that X-ray flares are produced by
internal shocks. Note that without the anisotropic correction, the
EIC light curve is higher and narrower and the peak EIC emission
is overestimated by one order of magnitude (see Fig.\ref{fig:WW}).

\begin{figure}
\begin{picture}(0,320)
\put(0,0){\includegraphics{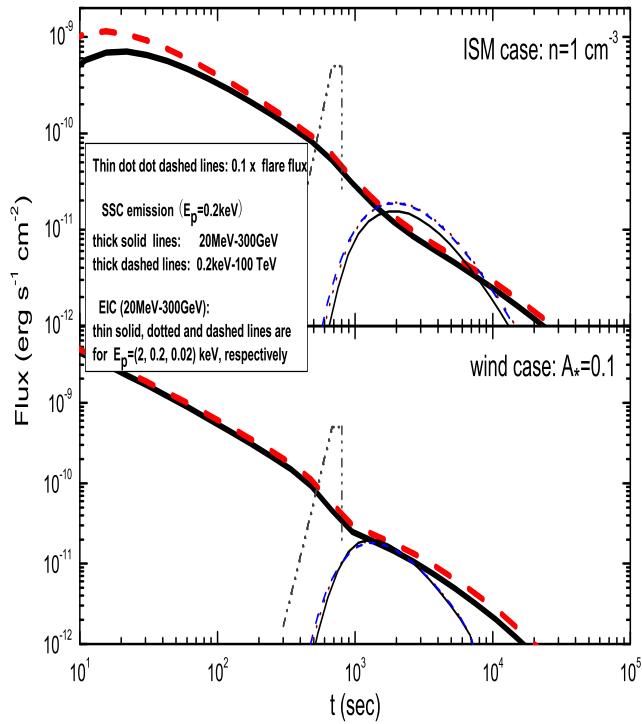}}
\end{picture}
\caption{High energy light curve arising from flare photons scattering
off the forward shock for the ISM/wind case (upper/lower panel),
respectively.  The parameters are the same as in
Figs. \ref{fig:EIC_dis} and \ref{fig:EIC_disb}, except for $E_{\rm p}$
which takes the values marked in the figure.} \label{fig:EIC_flux}
\end{figure}
\begin{figure}
\begin{picture}(0,200)
\put(0,0){\includegraphics{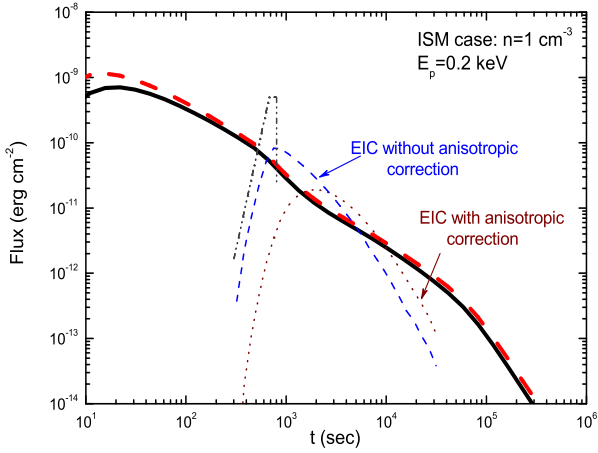}}
\end{picture}
\caption{The EIC emission with (the dotted line) and without (the
thin dashed line) anisotropic correction. Other lines and the
parameters are the same as in the upper panel of Figs.
\ref{fig:EIC_flux}, except for $E_{\rm p}$ that marked in the
figure.} \label{fig:WW}
\end{figure}

\begin{figure}
\begin{picture}(0,320)
\put(0,0){\includegraphics{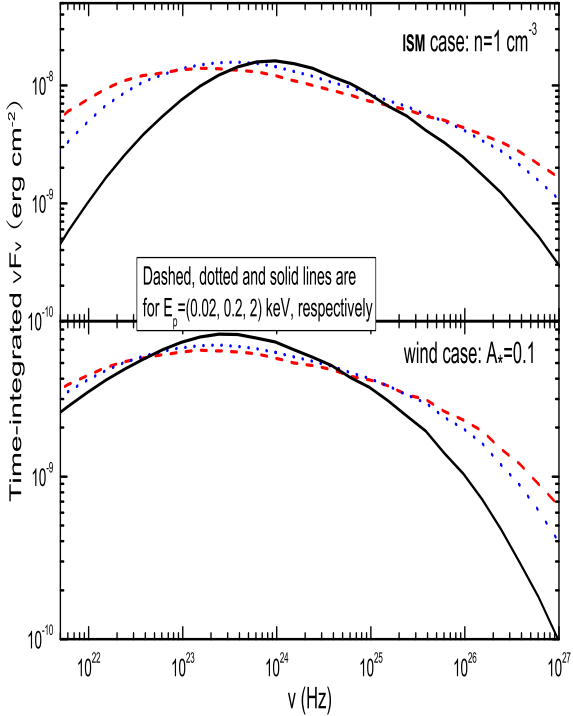}}
\end{picture}
\caption{The EIC spectrum resulting from upscattering of flare photons
by forward shock electrons.  The parameters used are the same as in
Fig. \ref{fig:EIC_dis} and Fig. \ref{fig:EIC_disb}, except for $E_{\rm
p}$ which takes the values marked in the figure.}
\label{fig:EIC_spectrum}
\end{figure}

For most soft X-ray flares, the peak emission energy seems to be
below 0.2 keV, i.e., they may be intrinsically far-UV flares. The
up-scattering of the far-UV photons in the external blast wave
results in strong sub-GeV emission. Fig. \ref{fig:EIC_spectrum}
depicts the resulting EIC spectrum (time integral) for different
values of $E_{\rm p}= (0.02,~0.2,~2)$ keV---other parameters,
including the luminosity of the flare in the 0.2-10keV band are
taken to be the same. For a far-UV flare ($E_{\rm p}\leq 0.2$ keV)
the seed photons are much more numerous than for an X-ray flare.
Consequently the resulting sub-GeV photons are much more numerous
than those resulting from a keV flare. Therefore the EIC emission
following a UV flare will be easier to detect.

\section{Detectability of high energy emission in the afterglow}\label{sec:det}

We turn now to the key question: Are the GeV to TeV high energy
signals predicted by our models observable with current or soon to be
commissioned detectors?

Using the calculated high energy spectrum $F_{\nu}(t)$ as a function
of time for any given model, we can estimate the total number, $N_{\rm
det}$, of detectable high energy photons,
\begin{equation}
N_{\rm det}=\int^{t_{\rm E}}_{t_{\rm I}} \int^{\nu_{\rm
u}}_{\nu_{\rm d}} {F_{\nu}(t) \over h\nu}S_{\rm det}(\nu) dt d\nu,
\label{eq:N_det}
\end{equation}
where $t_{\rm I,E}$ are the times when the observations begin and end,
respectively, $h\nu_{\rm d}-h\nu_{\rm u}$ is the energy range of the
detector, and $S_{\rm det}(\nu)$ is the effective area of the detector
as a function of $\nu$. For LAT onboard GLAST, we approximate $S_{\rm
det}(\nu)$ as (see http://www-glast.slac.stanford.edu
/software/IS/glast$_{-}$lat$_{-}$performance.htm):
\begin{equation} S_{\rm
det}(\nu) =
\left\{%
\begin{array}{ll}
     500~{\rm cm^2}~(h\nu/20{\rm MeV}),
&   \hbox{for $h \nu< 400{\rm MeV}$;} \\
      10^4{\rm cm^2},
& \hbox{for $h\nu \geq 400{\rm MeV}$.} \\
\end{array}%
\right. \label{eq:S_det}
\end{equation}

We consider first the high energy SSC emission in the afterglow,
which we estimated in sec. 4.  For the models presented in Fig.
\ref{fig:SSC1a} -- Fig.  \ref{fig:SSC_evo}, we use $t_{\rm I}\sim
100$ s; at earlier time the high energy emission may be dominated
by the synchrotron and/or SSC emission of the internal shocks
\cite{GZ07}. We choose an upper limit of $t_{\rm E}\sim 4\times
10^4$ s; after this time the SSC emission is usually too low to be
of interest.

\begin{figure}
\begin{picture}(0,200)
\put(0,0){\includegraphics{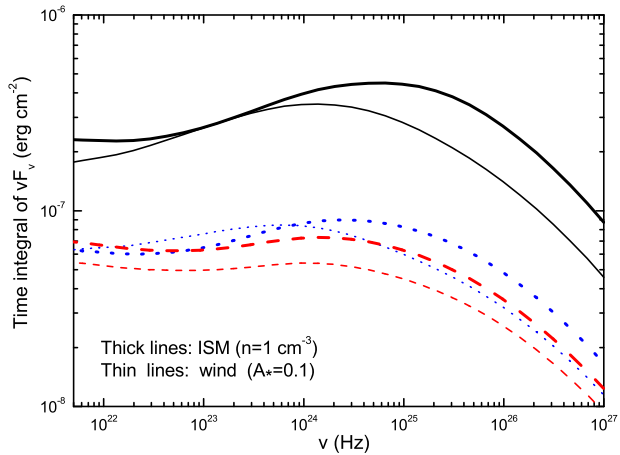}}
\end{picture}
\caption{The integral of $\nu F_\nu$ in the time interval
$100-4\times10^4$ s after the burst. The solid, dotted and dashed
lines correspond to the standard afterglow model (Figs.
\ref{fig:SSC1a} and \ref{fig:SSC1b}), the energy injection model
(Fig. \ref{fig:SSC_inj}) and the increasing $\epsilon_e$ model (Fig.
\ref{fig:SSC_evo}), respectively.} \label{fig:nuFnu}
\end{figure}

Figure \ref{fig:nuFnu} shows the integrated flux expected for the
various SSC scenarios discussed in sec. 4, and Table \ref{tab:1}
summarizes the expected number of photons that would be detected
by LAT from a burst with standard parameters (see Figs.
\ref{fig:SSC1a} - \ref{fig:SSC_evo}) at $z=1$. Typically, one
expects to detect a few photons above 20 MeV and  very few high
energy photons above 100 GeV.

\begin{table*}
   \begin{center}
     \caption{\label{tab:1} \small Expected signal for GLAST from SSC emission
     of a GRB forward shock, where the absorption of the very high energy photons by IR background
     is ignored. These values we calculate for a typical burst with
     $E_{\rm k}\sim 10^{53}{\rm erg}$
     (at the end of the X-ray shallow decline) and $z=1$, correspondingly to
     the burst with a
     $\gamma-$ray fluence of $\sim 10^{-5}~{\rm erg~cm^{-2}}$.}
     \begin{minipage}{18.5cm}
       \begin{tabular}{ccccc} \hline\hline
          & $N_{\rm det}(>20{\rm MeV})$ & $N_{\rm
det}(>100{\rm GeV})$
          \\ \hline

standard afterglow: ISM (Fig.\ref{fig:SSC1a})  & $\sim 13$ & $\sim
0.015$   \\
standard afterglow: wind (Fig.\ref{fig:SSC1b})  & $\sim 11$ &
$\sim
0.008$   \\
energy injection: ISM (Fig.\ref{fig:SSC_inj})  & $\sim 3.3$ &
$\sim
0.003$   \\
energy injection: wind (Fig.\ref{fig:SSC_inj})  & $\sim 3.6$ &
$\sim
0.002$   \\
time increasing $\epsilon_e$: ISM (Fig.\ref{fig:SSC_evo})  & $\sim
3.3$ & $\sim
0.002$   \\
time increasing $\epsilon_e$: wind (Fig.\ref{fig:SSC_evo})  &
$\sim 2.6$ & $\sim
0.001$   \\
\hline
       \end{tabular}
\vspace*{-0.4cm}
     \end{minipage}
   \end{center}
 \end{table*}

Not surprisingly, the modified afterglow models that account for
the shallow X-ray light curve in phase-II give fewer counts than
the standard afterglow model.  The reduced X-ray flux in these
models (needed to explain the shallow light curve) causes a
corresponding reduction in the high energy flux.  However, we
still expect a weak detection by GLAST. Such weak signals, of
course, cannot play an important role on distinguishing between
the different models. But for some extremely bright events, e.g.,
GRB 940217, the high energy observation may pose a tight
constraint on the underlying physical process (e.g., Wei \& Fan
2007).

Considering next the high energy emission associated with flares,
the time-integrated $\nu F_\nu$ of the high energy EIC component
is shown in Fig. \ref{fig:EIC_spectrum}. In the case of a uniform
external ISM, for $E_p=(0.02,~0.2,~2)$ keV, $N_{\rm det}(>20{\rm
MeV})=(0.6,~0.5,~0.2)$ and $N_{\rm det}(>100{\rm
GeV})=(3.2,~3.2,~2.6)\times 10^{-4}$, respectively. The EIC high
energy afterglow component is more easily detected if the flare
has a significant UV component. Note that Fan \& Piran (2006b)
used a larger effective detection area ($S_{\rm det}\sim 8000{\rm
cm^2}$ in the energy range of $20{\rm MeV}-300{\rm GeV}$). This
overestimates $S_{\rm det}$ for $h\nu<100$MeV where most of the
up-scattered photons are expected (see eq. \ref{eq:S_det}). In the
case of an external shock in a stellar wind we have for
$E_p=(0.02,~0.2,~2)$ keV, $N_{\rm det}(>20{\rm
MeV})=(0.3,~0.3,~0.2)$ and $N_{\rm det}(>100{\rm
GeV})=(1.7,~1.5,~1.0)\times10^{-4}$, respectively.  Now the
scattered far-UV photons are in the sub-MeV band (Fan \& Piran
2006b). Therefore, the far-UV component does not increase the
detected signal.

As long as the flare outflow is just weakly or even not magnetized
and $R_{\rm prompt}\geq 10^{14}$ cm, the GeV SSC emission fluence
is expected to be comparable to the fluence of the keV flare,
typically $10^{-7}-10^{-6}~{\rm erg~cm^{-2}}$. With such a
fluence, the GeV flashes (SSC emission) accompanying bright flares
may be detectable by GLAST.  If the flare is produced by a late
internal shock we expect that the typical SSC photon energy is
about 300 MeV. At this energy a fluence $\sim 2.4\times 10^
{-3}/S_{\rm det} \sim 3\times 10^{- 7}~{\rm erg ~ cm^{-2}}$
corresponds to a detection of 5 photons. So the SSC emission of a
very bright X-ray flare with a fluence $\sim 10^{-6} ~{\rm erg ~
cm^{-2}}$ should be detected, provided that the Compton parameter
is unity or larger. If the flare is produced by a refreshed shock
the typical photon energy would be higher, up to tens of GeV. The
number of these high energy photons would then be much smaller
than in the case of late internal shocks. As a result, it might
not be detectable by GLAST.

So far we have focused on the detectability of high energy emission
by GLAST.  However, there are also other detectors.
MAGIC\footnote{http://wwwmagic.mppmu.mpg.de/},
Whipple\footnote{http://veritas.sao.arizona.edu/old/VERITAS$_{-}$whipple.html}
and H.E.S.S. \footnote{http://www.mpi-hd.mpg.de /hfm/HESS/HESS.html}
are high energy telescopes operating at energies above $100$ GeV.
These Cerenkov detectors have very large effective areas $\sim
10^4-10^5{\rm m^2}$. The expected fluxes of very high energy
($>100$GeV) photons from bursts at $z \approx 1 $ should correspond
to the detection of $10-100$ photons. However, this estimate ignores
the absorption of the high energy photons by the IR-background
\cite{Nikoshov62}. Given that the optical depth for a $100$ GeV
photon from $z=3$ is $> 5$ \cite{Primack05}, we expect that for most
{\it Swift} bursts with a typical $z\sim 2.8$ the number of
detectable $>100$GeV photons will be negligible. Our results are
thus largely consistent with the null detection of the MAGIC
telescope \cite{MAGIC1}. Whipple \cite{Horan07} observed the
$>400$GeV afterglow emission of a few GRBs with $z\leq 1$, and in
particular GRB 030329, a nearby long burst. However, the earliest
observation was carried out $\sim 64.55$ hours after the trigger of
GRB 030329 when the expected very high energy SSC afterglow emission
is quite low (see also Xue et al. 2008). As the optical depth for IR
absorption increases strongly with energy and at low redshifts
linearly with $z$, we expect a detection of $100$ GeV or lower
energy photons from nearby strong bursts, provided the observations
begin very early and last for several hours. Such nearby bursts are,
of course, very rare but they do exist and high energy observatories
should focus on them.


\section{Summary and Discussion}\label{sec:Summary}
\begin{figure}
\begin{picture}(0,250)
\put(0,0){\includegraphics{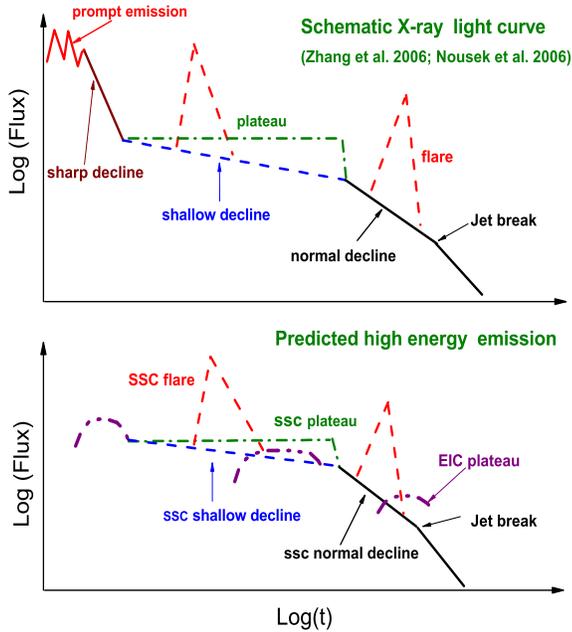}}
\end{picture}
\caption{Summary of the results. The expected high energy
afterglow signatures are shown in the lower panel, corresponding
to the schematic X-ray afterglow light curve shown in the upper
panel.  Note that the EIC emission light curve could be
outshined by the SSC emission of the forward shock.
 The SSC emission of the X-ray flares might be weak if the emission
 region is significantly magnetized or $R_{\rm flare}$ is much smaller than
 $10^{14}$ cm.} \label{fig:summary}
\end{figure}

Very high energy inverse Compton emission is an integral part of the
current afterglow model
\cite{MR94,Dermer00,Sari01,ZM01b,GM07,Galli06b,yu07,B05,Fan05b,Wang06,FP06b}.
We have calculated the high energy emission in different models of
GRB afterglows, including the SSC component of the forward shock,
the SSC component of the electrons producing X-ray flares and the
EIC component of flare photons upscattered by relativistic electrons
in the forward shock.  Our predicted high energy light curves are
summarized schematically in the lower panel of Fig.
\ref{fig:summary}.

High energy SSC emission in the energy range 20 MeV -- 300 GeV from
bright bursts should lead to a detectable signal of several ($\approx
10$) photons by the LAT onboard GLAST.  Higher energy telescopes such
as MAGIC, Whipple, H.E.S.S and Kangooroo working in the energy range
$>100$GeV, could detect strong signals (few hundred photons) from
nearby bursts at around the lower energy limit of these detectors
($\sim 100$GeV). Signals from more distant bursts will be absorbed by
the IR background.  The flux from a $z>1$ burst will usually be too
low to be detected.

Strong GeV SSC emission simultaneous with keV flare photons is
possible if the emitting region is not highly magnetized
\footnote{GeV flat segments followed by a sudden drop are also
expected to accompany the X-ray plateaus detected in GRB 060607A
\cite{Molinari07,JF07} and GRB 070110 \cite{Troja07}.}. The EIC
component of the flare, on the other hand, will be extended and will
last up to ten times as long as the X-ray flare (see Fig.
\ref{fig:EIC_flux})\footnote{Interesting EIC emission accompanying
Phase-I is also expected.}. This is because, in the EIC process, the
duration of the high energy emission is affected by the spherical
curvature of the blast wave and is mainly extended by the highly
anisotropic radiation of the up-scattered photons (see
Fig.\ref{fig:WW}). Unfortunately, a significant detection is likely
only if the SSC emission of the forward shock is very weak. A high
energy detection could be used to probe the spectrum of the low
energy flare and in particular the possible existence of a far UV
component. These signatures of the high energy flare are independent
of the density profile of the external medium.

A detection of a high energy component, {\it in principle}, will
enable us to test current models of GRBs and their afterglows.  A
detailed comparison of the high energy and low energy light
curves, in particular during the shallow decline phase, might even
enable us to discriminate between different modifications of the
standard afterglow model. However, given the small number of
expected high energy photons, it is unlikely that we can achieve
this goal with GLAST.

It should be noted that the two modifications to the standard model
that we considered in this paper, viz., extended energy injection
and time-evolving $\epsilon_e$, both predict achromatic behavior
such that there should be a shallow light curve in the optical band
simultaneously with the shallow X-ray light curve.  However, as
noted in Fig. 1, this is not always seen.  Thus, it is possible that
none of our models gives a correct description of afterglow physics.
Model that suggest flattening of the light curve due to emergence of
an X-ray SSC component (see Fig. \ref{fig:XRT}) that would arise in a
very dense medium may explain the X-ray flattening but this would
involve significant variability in the observed X-ray spectrum.

One intriguing possibility is that the standard afterglow model,
without energy injection or varying $\epsilon_e$, is indeed the
correct model, but the X-ray emission is suppressed during phase-II
because of some radiation physics that we have not yet understood.
This would explain why the optical light curve shows no shallow
phase-II segment or a break from phase-II to phase-III.  What kind of
high energy emission do we then expect?  In the absence of a real
model, we cannot say anything definite.  It is possible that the high
energy light curve would follow the predictions of the standard model.
Perhaps the high energy emission may even be enhanced because the
missing X-ray emission is radiated in this band.  These are pure
speculations, but they are worth keeping in mind.  It is very
important to carry out high energy observations of GRB afterglows,
independent of model expectations, because the signal may in the end
turn out to be stronger than anything predicted.

\section*{Acknowledgments}
We thank the second referee for detailed comments. YZF thanks X. Y.
Wang and Z. Li for discussions. This work is supported by US-Israel
BSF. TP acknowledges the support of Schwartzmann University Chair.
YZF and DMW are also supported by the National Science Foundation
(grant 10673034) of China and by a special grant of Chinese Academy
of Sciences.

\end{document}